\documentclass[aip,apl,reprint,floatfix,citeautoscript,longbibliography,superscriptaddress]{revtex4-1}
\usepackage{graphicx}
\usepackage{bm,amsmath,amssymb,mathrsfs,dcolumn}
\usepackage[usenames]{color}

\usepackage{subfigure}
\usepackage{ulem}
\begin{document}
\title{Magnon-photon coupling in antiferromagnets}
\author{H. Y. Yuan}
\affiliation{Department of Physics, Southern University of Science and Technology of China,
Shenzhen 518055, China}
\author{X. R. Wang}
\affiliation{Department of Physics, The Hong Kong University of
Science and Technology, Clear Water Bay, Kowloon, Hong Kong}
\affiliation{HKUST Shenzhen Research Institute, Shenzhen 518057, China}
%
\begin{abstract}
Magnon-photon coupling in antiferromagnets has many attractive features
that do not exist in ferro- or ferrimagnets. We show quantum-mechanically
that, in the absence of an external field, one of the two degenerated
spin wave bands couples with photons while the other does not.
The photon mode anticrosses with the coupled spin waves when their
frequencies are close to each other.
Similar to its ferromagnetic counterpart, the magnon-photon coupling
strength is proportional to the square root of number
of spins $\sqrt{N}$ in antiferromagnets. An external field removes the
spin wave degeneracy and both spin wave bands couple to the photons,
resulting in two anticrossings between the magnons and photons.
Two transmission peaks were observed near the anticrossing frequency.
The maximum damping that allows clear discrimination of
the two transmission peaks is proportional to $\sqrt{N}$ and it's well
below the damping of antiferromagnetic insulators. Therefore the strong
magnon-photon coupling can be realized in antiferromagnets and the
coherent information transfer between the photons and magnons is possible.
\end{abstract}
%
\maketitle
%

Information transfer between different information carriers is
an important topic in information science and technology.
This transfer is possible when strong coupling exists among different
information carriers. Strong coupling has already been realized between
photons and various excitations of condensed matter including electrons,
phonons, \cite{Tolpygo1950,Kun1951} plasmons \cite{Ritchie1957,
Barnes2003,Berini2011}, superconductor qubits \cite{Wallraff2004}, excitons
in a quantum well \cite{Dufferwiel2015} and magnons \cite{Soykal2010,
Huebl2013,Cao2015}. Among all of the excitations, magnons, which are
excitations of the magnetization of a magnet, are promising information
carriers in spintronics because of their low energy consumption, long
coherent distance/time, nanometer-scale wavelength, and useful information
processing frequency ranging from gigahertz (GHz) to terahertz (THz).
Furthermore, magnons can also be a control knob of magnetization dynamics
\cite{yanpeng2011,hubin2013,xiansi2012}, and magnon bands of a magnet
can be well controlled by either magnetic field or electric current.
The electric field $\mathbf{E}$ and magnetic inductance $\mathbf{B}$ in a
microcavity of volume $V$ can be sufficient strong even with only one or
a few photons of frequency $\nu$ ($|\mathbf{E}|$,
$|\mathbf{B}| \propto \sqrt{h\nu/V}$).
Therefore, the coupling between the microcavity photons and the magnons of
nanomagnets have received particular attention in recent years.
Moreover, similar to the cavity quantum electrodynamics \cite{Walther2006}
which deals with coupling between photons and atoms in a cavity and
provides a useful platform for studying quantum phenomenon and for various
applications in micro laser and photon bandgap structure, cavity magnonics
is also a promising arena for investigating magnons at the quantum level and
for manipulating information transfer between single photon and single magnon.

The theoretical demonstration of a possible coupling of
a ferro-/ferrimagnet to light was provided in 2013 \cite{Soykal2010}.
The coupling strength is proportional to the square root of the number of
spins $\sqrt{N}$ and the coupling energy could be as big as $\sim 100\
\mu eV$ in a cavity of $\sim 1$ mm and resonance frequency $\sim 200$ GHz.
The prediction was experimentally confirmed by placing a yttrium iron
garnet (YIG) particle in a microwave cavity of high quality factor.
\cite{Tabuchi2014, Zhang2014, Bai2015, Huebl2013}
Many applications based on these results have been proposed, including
the generation and characterization of squeezed states through the
interaction between magnons and superconducting qubits via microwave
cavity photons \cite{Tabuchi2014} and coherent information transfer
between magnons and photons \cite{Zhang2014}. The information can be
transmitted and read out electrically in the hybrid architecture under a
strong magnon-photon coupling.
\cite{Flaig2016}

Antiferromagnets (AFM) have many useful properties in comparison
with ferromagnetic materials such as better stability against the
external field perturbations and negligible cross talking with the
neighboring AFM elements because of the absence of stray fields.
The AFM dynamics is typically of the order of THz, much faster than
the order of GHz for ferromagnets. Because of these superb properties,
various aspects of antiferromagnetic spintronics have attracted
significant interests in the last few years including domain wall
motion, skyrmions, magnetoresistence, magnetic switching, spin
pumping, spin current transport and so on. \cite{Jungwirth2016}
However, only few works based on the classical electrodynamics were
reported on the magnon-photon coupling \cite{Manohar1972,Bose1975}
in AFM so far. In order to have a better understanding of the
magnon-photon coupling in AFM, we would like to study the issue
at the quantum level. In this letter, we demonstrate quantum
mechanically the existence of magnon-polariton in an AFM and show
that there exists a dark mode and a bright mode in the strong
coupling regime. Antiferromagnetic insulators with low damping are
promising candidates to realize strong magnon-photon coupling.

\begin{figure}
\centering
\includegraphics[width=0.4\textwidth]{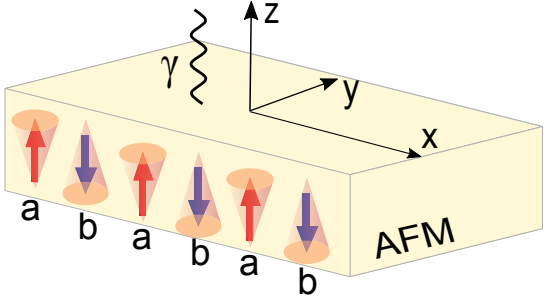}\\
\caption{(color online) Sketch of an antiferromagnetic N\'{e}el
state interacted with light. The $x,y,z$ axes are respectively
along the length, width, and thickness directions of the AFM strip.
The red and blue arrows, respectively pointing to the $+z$ and $-z$
orientations, represent spins on sublattices $a$ and $b$.}
\label{fig1}
\end{figure}

We consider a two sublattice antiferromagnet whose spins on the sublattices
($a$ and $b$) align in the opposite directions along $\pm z$-axis as
shown in Fig. \ref{fig1}. The Hamiltonian of the AFM coupled with light
through its magnetic field is
\begin{equation}
\begin{aligned}
&H=H_{\mathrm{AFM}} + H_{\mathrm{ph}} + H_{\mathrm{int}},\\
&H_{\mathrm{AFM}}=J \sum_{l,\delta} ( \mathbf{S}_l^a \cdot \mathbf{S}_{l+\delta}^b
+ \mathbf{S}_l^b \cdot \mathbf{S}_{l+\delta}^a ) \\
&\ -\sum_l (\mathbf{H}_0 + \mathbf{H}_a) \cdot \mathbf{S}_l^{a}
- \sum_l (\mathbf{H}_0 - \mathbf{H}_a) \cdot \mathbf{S}_{l + \delta}^{b}\\
&H_{\mathrm{ph}}=\frac{1}{2}\int \left ( \epsilon_0 \mathbf{E}^2
+ \frac{1}{\mu_0}\mathbf{B}^2 \right ) \mathrm{d}x\mathrm{d}y\mathrm{d}z\\
&H_{\mathrm{int}}=-\sum_{l,\alpha=a,b} \mathbf{S}_l^\alpha \cdot \mathbf{H}_f
\end{aligned}
\label{hamiltonian}
\end{equation}
where $H_{\mathrm{AFM}}$, $H_{\mathrm{ph}}$, $H_{\mathrm{int}}$ are
respectively the Hamiltonian for AFM, photon and their interaction.
$J$ ($>0$) is the exchange constant, $\mathbf{S}_l^a$ and $\mathbf{S}_l^b$
are the spins on sites $l$ of sublattices $a$ and $b$ respectively.
$\delta$ denotes the displacement of two nearest spins.
$\mathbf{H}_0$ is the external magnetic field and $\mathbf{H}_a$ is
the anisotropy field. $\mathbf{E}$ and $\mathbf{B}$ are the electric
field and magnetic inductance of the electromagnetic (EM) wave and
$\mathbf{H}_f$ is the corresponding magnetic field, $\epsilon_0$ and
$\mu_0$ are vacuum permittivity and susceptibility, respectively.

Using the Holstein-Primakoff transformation \cite{Holstein1940},
$H_{\mathrm{AFM}}$ in the momentum space can be written as
\begin{equation}
\begin{aligned}
H_{\mathrm{AFM}}&=H_{\mathrm{ex}} \sum_q \left [ \gamma_q (a_q^\dagger
b_q^\dagger + a_q b_q) + (a_q^\dagger a_q +b_q^\dagger b_q) \right ] \\
&+\sum_q  \left [(H_a+H_0) a_q^\dagger a_q+(H_a- H_0)b_q^\dagger b_q
\right ],\\
\end{aligned}\label{qm}
\end{equation}
where $H_{\mathrm{ex}} = 2JSz$, $z$ and $\gamma_q$ are respectively
the coordination number and the structure factor of the lattice.
The EM wave could be quantized through the standard procedures
$H_{\mathrm{ph}} = \hbar \sum_q \omega_q \left ( c_q^\dagger c_q
+ \frac{1}{2} \right )$ and the interaction term is
$H_\mathrm{int} = \hbar \sum_q g_c \left ( c_qa_q + c_q^\dagger a_q^
\dagger+ c_q b_q^\dagger + c_q^\dagger b_q\right)$ for a circularly polarized
wave, where $g_c = \sqrt{\mu_0 \omega_q S N/ 2 \hbar V}$, $\hbar$, $N$,
$V$ and $\omega_q$ are respectively the Planck constant, the number of spins
on each sublattice, the volume of the cavity, and the photon frequency.
The photon dispersion relation is linear
$\omega_q=c|\mathbf{q}|$, where $c$ is the speed of light.
$a_q^+, a_q$, $b_q^+, b_q$ and $c_q^+, c_q$ are creation and annihilation
operators of magnons and photons, respectively, and they satisfy the
commutation relations for bosons.

The Hamiltonian (\ref{qm}) does not conserve the magnon number, and can
be diagonalized by the Bogoliubov transformation,
\begin{equation}
a_q =u_q \alpha_q +v_q\beta_q^\dagger, b_q=u_q \beta_q+ v_q
\alpha_q^\dagger, \\
\label{bt}
\end{equation}
where $u_q=\sqrt{(\Delta_q -1)/2}$, $v_q =\sqrt{(\Delta_q +1)/2}$,
and $\Delta_q = 1/\sqrt{1-(H_{ex}\gamma_q/(H_{ex}+H_a))^2}$.
In terms of the boson oeprators $\alpha_q, \alpha_q^\dagger, \beta_q, \beta_q^\dagger$,
$H_{\mathrm{AFM}}$ reads
\begin{equation}
H_{\mathrm{AFM}} = \sum_q \hbar  \omega_q^- \alpha_q^\dagger
\alpha_q+ \hbar \omega_q^+\beta_q^\dagger \beta_q,
\end{equation}
where
\begin{equation}
\begin{aligned}
\omega_q^\pm = \pm \gamma H + \gamma \sqrt{ H^2_{\mathrm{sp}}
+ H_{\mathrm{ex}}^2 (1-\gamma_q^2)}
\end{aligned}
\end{equation}
is magnon dispersion relation of an AFM. $\gamma$ is gyromagnetic
ratio and
$H_{\mathrm{sp}}=\sqrt{ H_\mathrm{a}(H_\mathrm{a} + 2H_{\mathrm{ex}})}$
is the spin-flop transition field. Under the transformation
of Eq. (\ref{bt}), the interaction Hamiltonian is
\begin{equation}
H_{\mathrm{int}} = \hbar \sum_q g_c(u_q+v_q) (c_q \alpha_q
+ c_q^\dagger \alpha_q^\dagger
+ c_q\beta_q^\dagger  + c_q^\dagger \beta_q).
\end{equation}

Because the slope of the photon dispersion relation is much more
steep than that of the magnon, the photon can only interact
strongly with the magnons around the Gamma point ($q=0$). For simplicity,
we set $q=0$ and the sum in the $H_{\mathrm{AFM}}$ is removed.
To obtain the eigen-modes of the coupled system, we define
$\Psi = (\alpha_q, \beta_q^\dagger, c_q^\dagger)^\dagger$ and write
the Hamiltonian in the matrix form
$H=\hbar \Psi ^\dagger \mathbf{M}\Psi$ with

\begin{equation}
\mathbf{M}=\left (
\begin{array}{ccc}
    \omega^- & 0 & \lambda/2 \\
    0 & \omega^+ & \lambda/2 \\
    \lambda/2 & \lambda/2 & \omega_c
  \end{array}\right ).
\end{equation}
where $\lambda=2g_c(u+v)=2g_c(u_{q=0} + v_{q=0}),\omega^\pm = \omega_{q=0}^\pm$
and $\omega_c$ is the photon frequency.

The eigen-equation of $\mathbf{M}$ reads
\begin{equation*}
\begin{aligned}
4 \omega^3 - 4 (\omega^+ + \omega^- + \omega_c) \omega^2
+ \lambda^2 (\omega^+ + \omega^- )-4\omega^+ \omega^- \omega_c\\
+2 (-\lambda^2 + 2\omega^+ \omega^- + 2\omega^+ \omega_c
+ 2\omega^- \omega_c)\omega =0.\\
\end{aligned}
\end{equation*}

\begin{figure}
\centering
\includegraphics[width=0.45\textwidth]{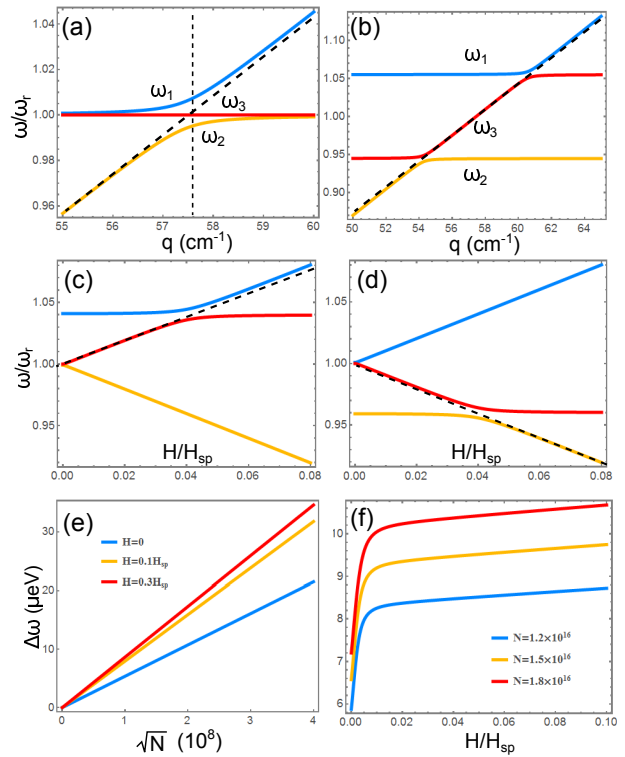}\\
\caption{(color online) Magnon-polariton spectrum for $H=0$ (a)
and $H=0.15H_{\mathrm{sp}}$ (b), respectively.
$N=1.56 \times 10^6$, $H_a=0.016H_{ex}$.
The dashed lines refer to the photon mode $\omega = cq$.
(c-d) The frequency of the magnon-photon system as a function of
external field at given photon frequency $\omega_c = 1.04\omega_r$
and $0.96\omega_r$, respectively. when the external fields is adjusted.
(e) The band gap of the coupled spectrum $\Delta \omega$ as a function
of the number of spins under fields $H=0$, $H=0.1H_{sp}$ and
$H=0.3H_{\mathrm{sp}}$, respectively. (f) $\Delta \omega$ as a function
of the external fields for  $N=1.2 \times 10^6$, $N=1.5 \times 10^6$
and $N=1.8 \times 10^6$, respectively. Other parameters are
$H_{\mathrm{ex}} = 54$ T, $S=1$, $V=1\ \mathrm{mm^3}$, $a=0.4$ nm.}
\label{fig2}
\end{figure}

In the absence of an external field, this cubic equation has
analytical solutions
\begin{equation}
\begin{aligned}
&\omega_{1,2}= \frac{1}{2} \left [ \omega_r + \omega_c \pm
\sqrt{ (\omega_r - \omega_c)^2 + 2\lambda^2} \right ], \\
&\omega_3 =\omega_r=\gamma H_{\mathrm{sp}}.
\end{aligned}
\label{zero_h}
\end{equation}
The typical dispersion relation is shown in Fig. \ref{fig2}a.  Obviously,
one magnon band is left unchanged ($\omega_3$, red line) and the other
band is coupled with the photon mode and anticrosses with each other
($\omega_1$ and $\omega_2$, blue and yellow lines). Therefore, one of
the degenerated magnon band at $H=0$ is a dark mode that doesn't
interact with the photons while the other is a bright mode and interacts
with the photon. For very small and very large wavevector $q$,
the linear dispersion are mainly from the photons (dashed line).
Only near the wavevector $q = \omega_r$/c, where the photon frequency
equals magnon frequency, the anticrossing feature becomes pronounced.

When the external field is non-zero, the double degeneracy of the
magnon modes are removed with an energy split proportional to $2H$.
Both magnon bands are coupled with the cavity photon, but the two
anticrossings appear at two different $q$, which is determined by
$q = \omega^\pm /c$, as shown in Fig. \ref{fig2}b. On the other hand,
for a fixed photon frequency of $\omega_c$, strong coupling occurs
by adjusting the external field $H$ so that $\omega_\pm = \omega_c$.
Depending on the magnitude of the photon frequency, strong coupling
can be with either the ascending band $\omega_+$ or descending band
$\omega_-$, as shown in Fig. \ref{fig2}c and \ref{fig2}d, respectively.
Furthermore, to achieve a reliable information transfer between the
magnons and photons, it's important to know the coupling strength
between the magnons and photons. According to Eq. (\ref{zero_h}),
the frequency split of the two anticrossing modes at the resonance
is $\Delta \omega = \sqrt{2} \lambda = 2 \sqrt{2}g_c(u+v)$,
which is proportional to the coupling strength $g_c (u+v) $.
Thus we will express the coupling strength by $\Delta \omega$ below.
The coupling strength as a function of spin numbers $N$ is shown in
Fig. \ref{fig2}e. The coupling strength increases linearly with the
square root of $N$. For $N=2.0 \times 10^{16}, \ H=0.1H_{\mathrm{sp}}$,
the coupling is $11.3 \ \mu eV$. Figure \ref{fig2}f shows the
field-dependence of the coupling strength. The coupling strength first
increases sharply with the field and then approaches a constant value.

\begin{figure}
\centering
\includegraphics[width=0.45\textwidth]{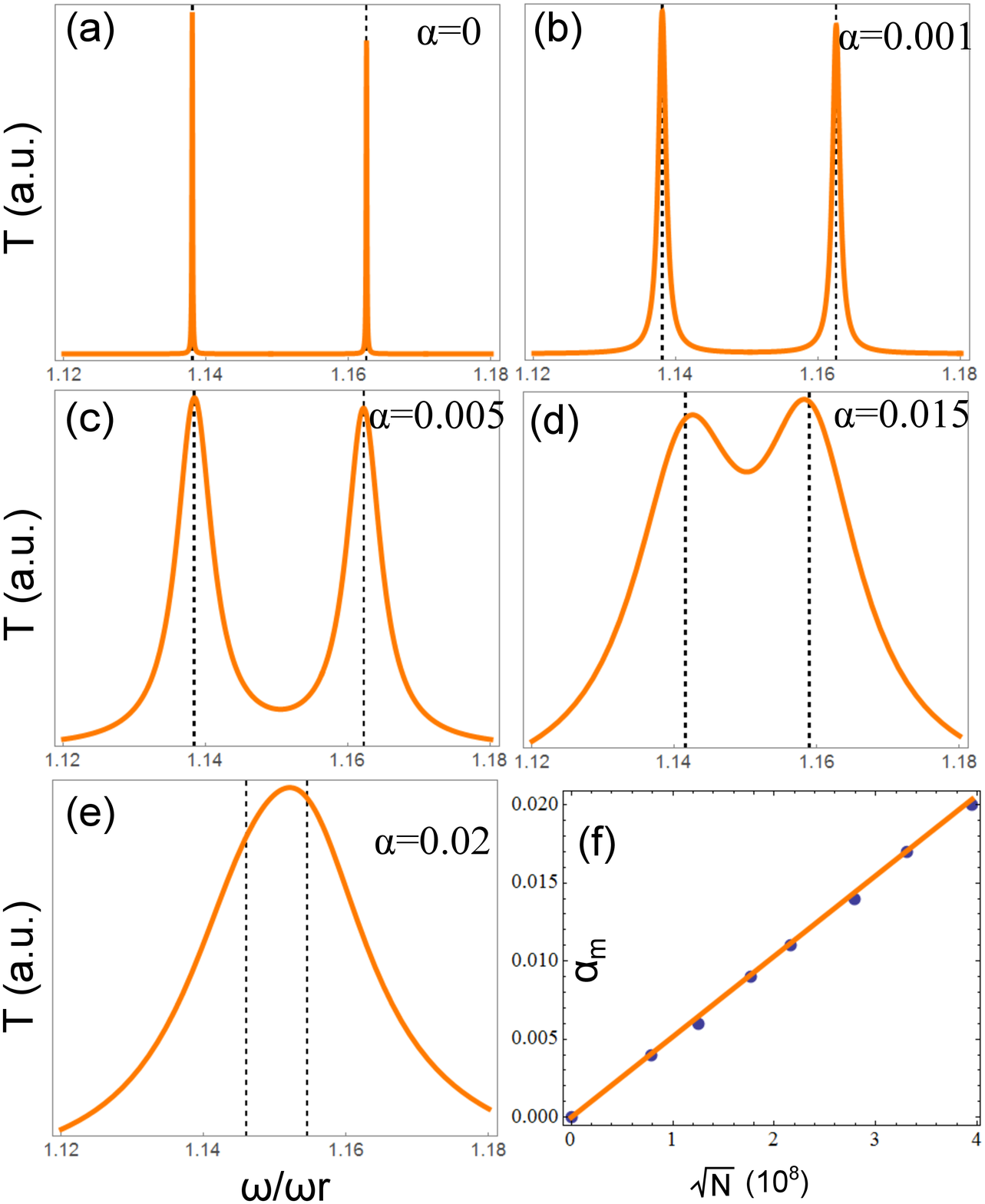}\\
\caption{(color online)
Transmission of incident wave at $\alpha=0$ (a), 0.001 (b),
0.005 (c), 0.015 (d) and 0.02 (e) for $H=0.15H_{\mathrm{sp}}$.
The vertical dashed lines indicate the positions of the coupled modes.
(f) The $\sqrt{N}-$dependence of maximum damping that can have two
distinguishable resonance peaks for $H=0.15H_{\mathrm{sp}}$.
The linear line is $\alpha_m = \Delta \omega/ \sqrt{2} \omega_c$,
where $\Delta \omega$ is proportional to $\sqrt{N}$.}
\label{fig3}
\end{figure}

The transmission of an incident EM wave is often measured in the experiments.
As argued in the previous publications \cite{Harder2016},
the transmission can be viewed as a scattering process, which is
well described by the Green function of a magnet-light system.
Suppose the eigenvectors of eigenvalues $\omega_{1,2,3}$ are
$\left|1\right>, \left|2\right>, \left|3\right>$, respectively.
Then the Green function in the diagonal basis is
\begin{equation}
G=\sum_{k=1,2,3}\frac{\left|k\right> \left<k\right | }{\omega -\omega_k
+i\epsilon}
\end{equation}
where $\epsilon$ is an arbitrary small positive number.
The transmission amplitude is the imaginary part of the Green function,
i.e. $\mathbf{T} (\omega) \propto - \mathrm{Im} (\mathbf{G} (\omega))$.
The transmission of an incident wave $\left | \varphi_0 \right >$
(eigen-mode of $c_q^\dagger c_q$) is
$T = \left < \varphi_0 \right | \mathbf{T} \left | \varphi_0 \right >$.
Figure \ref{fig3}a is the transmission near the photon frequency for
$H=0.15H_{sp}$ and $N=1.56 \times 10^7$.
Two transmission peaks center at the calculated eigen-frequency
(dashed lines), which demonstrates the strong magnon-photon coupling.
The $\delta$-function like transmission peaks are due to the absence of the
damping. In realistic case, the damping will broaden the peaks of the
Lorentzian curve. If the damping is large enough, the two Lorentzian
peaks will merge to a single peak and then the coupling modes cannot
be identified.

To quantitatively see the influence of damping on the transmission
spectrum, we first replace $\omega_r$ by $\omega_r - i\alpha\omega_r$
in the matrix $\mathbf{M}$, where $\alpha$ is the strength of damping,
then we calculate the complex eigenvalues and eigenvectors of $\mathbf{M}$
and use them to compute the imaginary part of the Green function
(transmission amplitude). Figure \ref{fig3}b-e is the frequency-dependence
of the transmission for $\alpha$ increasing from 0.001 to 0.02.
Indeed, (transmission) peak width increases, and peak height
decreases with the increase of $\alpha$. For the parameters used in
our calculations, two peaks become indistinguishable for damping larger
than 0.02. As the number of spins $N$ increases, the coupling strength
between magnons and photons increases, then the magnon-polariton is
more stable to resist the intrinsic damping of magnons.
Figure \ref{fig3}f shows the maximum damping $\alpha_m$ that
allows clear identification of the two coupled modes as a function
of $\sqrt{N}$ for $H=0.15H_{sp}$, which verifies the argument.
Quantitatively, the linewidth of the absorption curve should be
$c_0 \alpha \omega_c$, then the maximum damping could be derived as
$c_0 \alpha_m \omega_c = \Delta \omega$ i.e. $\alpha_m = \Delta \omega/(c_0 \omega_c)$.
The numerical data could be perfectly described by using $c_0=\sqrt{2}$,
as shown by the orange line in Fig. \ref{fig3}f.

Our results suggest that the key to realize the strong magnon-photon
coupling in AFMs is to use low damping materials.
The intrinsic damping of an antiferromagnetic metal is of the order of 0.5
according to the first principles calculation, which will be published
elsewhere.  Hence antiferromagnetic metals are not favorable for realizing
strong magnon-photon coupling. The damping of antiferromagnetic insulators
such as NiO can be as low as $2.1 \times 10^{-4}$ \cite{Kampfrath2011},
comparable to that of YIG. Thus, the strong coupling can be realized in
low damping antiferromagnetic insulators according to our results.
In terms of the detection of the coupling signal, we can either measure
the transmission spectrum or use electric detection method to
measure the voltage signal of a hybridized structure.
For the ferromagnetic case, the coupling strength between magnon and
microwave photons have been measured through the electrical detection
of spin pumping from the ferromagnetic layer \cite{Flaig2016}.
It was recently reported that spin pumping exists also at the interface
of AFM/normal metal \cite{Cheng2014}. In fact, AFM layer may even enhance the spin
pumping. Thus, electrical detection of the coupling
signal in AFM is also possible in the hybridized structures.
Furthermore, magnon modes in an AFM has already been experimentally
excited by using sub-THz technology \cite{Caspers2016}.

In conclusions, we have quantum-mechanically investigated the
magnon-photon coupling in an antiferromagnet. The coupling
strength is proportional to the square root of number of spins and
can be order of several $\mu eV$ to tens of $\mu eV$,
which could be observed in low damping AFM insulators.
In the absence of an external field, only
one magnon band is coupled with the cavity photon and anticrosses
with each other near the cavity frequency while the other does not.
External fields remove the double degeneracy in magnon bands and both
magnon bands couple to the cavity photon, resulting in two anticrossings.

HYY would like to thank Ke Xia, Zhe Yuan, Grigoryan Vahram and Meng Xiao
for helpful discussions.
XRW acknowledges the support from National Natural Science Foundation of China
(Grant No. 11374249) and Hong Kong RGC (Grant No. 163011151 and 16301816).


\begin{thebibliography}{}
\bibitem{Tolpygo1950} K. B. Tolpygo, Zh. Eksp. Teor. Fiz. \textbf{20}, 497 (1950).

\bibitem{Kun1951} K. Huang, Nature \textbf{167}, 779 (1951).

\bibitem{Ritchie1957} R. H. Ritchie, Phys. Rev. \textbf{106}, 874 (1957).

\bibitem{Barnes2003} W. L. Barnes, A. Dereux, and T. W. Ebbesen,
Nature \textbf{424}, 824 (2003).

\bibitem{Berini2011} P. Berini and I. De Leon, Nat. Photon.
\textbf{6}, 16 (2011).

\bibitem{Wallraff2004} A. Wallraff, D. I. Schuster, A. Blais,
L. Frunzio, R.-S. Huang, J. Majer, S. Kumar, S. M. Girvin,
and J. Schoekopf, Nature \textbf{431}, 162 (2004).

\bibitem{Dufferwiel2015} S. Dufferwiel, S. Schwarz, F. Withers,
A. A. P. Trichet, F. Li, M. Sich, O. Del Pozo-Zamudio, C. Clark,
A. Nalitov, D. D. Solnyshkov, G. Malpuech, K. S. Novoselov,
J. M. Smith, M.S. Skolnick, D. N. Krizhanovskii, and
A. I. Tartakovskii, Nat. Commun.
\textbf{6}, 8579 (2015).

\bibitem{Soykal2010} \"{O}. O. Soykal and M. E. Flatt\'{e},
Phys. Rev. Lett. \textbf{104}, 077202 (2010).

\bibitem{Huebl2013} H. Huebl, C. W. Zollitsch, J. Lotze,
F. Hocke, M. Greifenstein, A. Marx, R. Gross,
and S. T. B. Goennenwein, Phys. Rev. Lett. \textbf{111}, 127003 (2013).

\bibitem{Cao2015} Y. Cao, P. Yan, H. Huebl, S. T. B. Goennenwein,
and G. E. W. Bauer, Phys. Rev. B \textbf{91}, 094423 (2015).

\bibitem {yanpeng2011} P. Yan, X. S. Wang, and X. R. Wang, Phys. Rev. Lett.
\textbf{107}, 177207 (2011).
\bibitem{xiansi2012} X. S. Wang, P. Yan, Y. H. Shen, G. E. W. Bauer, and
X. R. Wang, Phys. Rev. Lett. \textbf{109}, 167209 (2012).
\bibitem {hubin2013} B. Hu and X. R. Wang, Phys. Rev. Lett.
\textbf{111}, 027205 (2013).

\bibitem{Walther2006} H. Walther, B. T. H. Varcoe, B. Englert,
and T. Becker, Rep. Prog. Phys. \textbf{69}, 1325 (2006).

\bibitem{Tabuchi2014} Y. Tabuchi, S. Ishino, T. Ishikawa,
R. Yamazaki, K. Usami, and Y. Nakamura, Phys. Rev. Lett.
\textbf{113}, 083603 (2014).

\bibitem{Zhang2014} X. Zhang, C.-L. Zou, L. Jiang, and H. X. Tang,
Phys. Rev. Lett. \textbf{113}, 156401 (2014).

\bibitem{Bai2015} L. Bai, M. Harder, Y. P. Chen, X. Fan, J. Q. Xiao,
and C. -M. Hu, Phys. Rev. Lett. \textbf{114}, 227201 (2015).

\bibitem{Flaig2016} H. Maier-Flaig, M. Harder, R. Gross,
H. Huebl, S. T. B. Geennenwein, arXiv:1601.05681v1

\bibitem{Jungwirth2016} T. Jungwirth, X. Marti, P. Wadley,
and J. Wunderlich, Nat. Nanotech. \textbf{11}, 231 (2016).

\bibitem{Manohar1972} C. Manohar and G. Venkataraman,
Phys. Rev. B \textbf{5}, 1993 (1972).

\bibitem{Bose1975} S. M. Bose, E-Ni. Foo, and M. A. Zuniga,
Phys. Rev. B \textbf{12}, 3855 (1975).

\bibitem{Holstein1940} T. Holstein and H. Primakoff,
Phys. Rev. \textbf{58}, 1098 (1940).

\bibitem{Harder2016} M. Harder, L. Bai, C. Match, J. Sirker,
and C. -M. Hu, arXiv:1601.06049v2.

\bibitem{Kampfrath2011} T. Kampfrath, A. Sell, G. Klatt,
A. Pashkin, S. M\"{a}hrlein, T. Dekorsy, M. Wolf, M. Fiebig,
A. Leitenstorfer, and R. Huber, Nat. Photon. \textbf{5}, 31 (2011).

\bibitem{Cheng2014} R. Cheng, J. Xiao, Q. Liu, and A. Brataas,
Phys. Rev. Lett. \textbf{113}£¬ 057601 (2014).

\bibitem{Caspers2016} C. Caspers, V. P. Gandhi, A. Magrez,
E. de Rijk, and Jean-Philippe Ansermet,
Appl. Phys. Lett. \textbf{108}, 241109 (2016).
\end{thebibliography}
\end{document}